\documentclass[twocolumn,prb,aps,showpacs]{revtex4}
\usepackage{epsf,latexsym,graphicx}
\usepackage{amssymb,amsmath}

\def\beeq {\begin{equation}}
\def\eneq {\end{equation}}
\def\ba {\begin{eqnarray}}
\def\ea {\end{eqnarray}}
\def\vk {\mathbf{k}}

\begin{document}

\title{Momentum Dependence of the Single-Particle Self-Energy and Fluctuation
Spectrum of Slightly Underdoped
Bi$_2$Sr$_2$CaCu$_2$O$_{8+\delta}$ from High-Resolution Laser
Angle-Resolved Photoemission}

\author{Jin Mo Bok }\affiliation{Department of Physics and Institute for
Basic Science Research, SungKyunKwan University, Suwon 440-746,
Korea.}

\author{Jae Hyun Yun}\affiliation{Department of Physics and Institute for
Basic Science Research, SungKyunKwan University, Suwon 440-746,
Korea.}

\author{Han-Yong Choi}
\affiliation{Department of Physics and Institute for Basic
Science Research, SungKyunKwan University, Suwon 440-746, Korea.}

\author{Wentao Zhang}
\affiliation{National Laboratory for Superconductivity, Beijing
National Laboratory for Condensed Matter Physics, Institute of
Physics, Chinese Academy of Sciences, Beijing 100190, China.}

\author{X. J. Zhou}
\affiliation{National Laboratory for Superconductivity, Beijing
National Laboratory for Condensed Matter Physics, Institute of
Physics, Chinese Academy of Sciences, Beijing 100190, China.}

\author{Chandra M. Varma}
\affiliation{Department of Physics and Astronomy, University of
California, Riverside, California 92521. \\}

\begin{abstract}

We deduce the normal state angle-resolved single-particle
self-energy $\Sigma(\theta, \omega)$ and the Eliashberg function
(i.e., the product of the fluctuation spectrum and its coupling to
fermions) $\alpha^2 F(\theta,\omega)$ for the high temperature
superconductor Bi$_2$Sr$_2$CaCu$_2$O$_{8+\delta}$ from the ultra
high resolution laser angle-resolved photoemission spectroscopy
(ARPES). The self-energy $\Sigma(\theta, \omega)$ at energy
$\omega$ along several cuts normal to the Fermi surface at the
tilt angles $\theta$ with respect to the nodal direction in a
slightly underdoped Bi$_2$Sr$_2$CaCu$_2$O$_{8+\delta}$ were
extracted by fitting the ARPES momentum distribution curves. Then,
using the extracted self-energy as the experimental input, the
$\alpha^2 F(\theta,\omega)$ is deduced by inverting the
Eliashberg equation employing the adaptive maximum entropy
method. Our principal new result is that the Eliashberg function
$\alpha^2F(\theta,\omega)$ collapse for all $\theta$ onto a
single function of $\omega$ up to the upper cut-off energy
despite the $\theta$ dependence of the self-energy. The in-plane
momentum anisotropy is therefore predominantly due to the
anisotropic band dispersion effects. The obtained Eliashberg
function has a small peak at $\omega\approx0.05$ eV and flattens
out above 0.1 eV up to the angle-dependent cut-off. It takes the
intrinsic cut-off of about 0.4 eV or the energy of the bottom of
the band with respect to the Fermi energy in the direction
$\theta$, whichever is lower. The angle independence of the
$\alpha^2 F(\theta,\omega)$ is consistent only with the
fluctuation spectra which have the short correlation length on
the scale the lattice constant. This implies among others that the
antiferromagnetic fluctuations may not be underlying physics of
the deduced fluctuation spectrum.

\end{abstract}

\pacs{PACS: } \keywords{}

\maketitle

\section{Introduction}

It is generally agreed that understanding the normal state
properties is prerequisite to the high temperature
superconductivity because the pairing instability is a normal
state Fermi surface instability. A number of studies have been
conducted to elucidate the normal state charge and spin dynamics
and momentum anisotropy of the high temperature
superconductors.\cite{Cho06science,Chubukov03,Basov05rmp,Hussey08jpcm}
The angle resolved photoemission spectroscopy (ARPES), owing to
its unique momentum and energy resolution, has been quite
powerful in uncovering the in-plane momentum anisotropy of the
quasi-particle (qp) dynamics of the
cuprates.\cite{Damascelli03rmp} Early ARPES measurements
\cite{Valla99science} on optimally doped (OP)
Bi$_2$Sr$_2$CaCu$_2$O$_{8+\delta}$ (Bi2212) and subsequent
measurements\cite{Bogdanov00prl,Kaminski00prl,Lanzara01nature,Zhou03nature}
along the nodal cut of $(0,0)-(\pi,\pi)$ direction showed the
marginal Fermi liquid (MFL) \cite{Varma89prl} behavior of $-Im
\Sigma(\omega)\propto \omega$ and found a kink around 0.06 eV in
the qp dispersion. Kaminski $et~al.$ extended these measurements
to off-nodal cuts to investigate the in-plane anisotropy of the qp
scattering rate $-Im \Sigma(\theta,\omega)$ around the Fermi
surface.\cite{Kaminski05prb} They reported that the functional
form of the scattering rate for under (UD) and optimally doped
sample can be written as $ a + b \omega$, where the elastic term
$a(\theta)$ is anisotropic in correlation with the pseudogap, and
the inelastic term, $b$, is isotropic around the Fermi surface;
the scattering rates become isotropic for the heavily overdoped
(OD) samples. On the other hand, Chang $et~al.$ reported that
both the elastic and inelastic terms are anisotropic, exhibiting
a minimum along the nodal direction.\cite{Chang08prb}

The in-plane anisotropy may also be probed by the angle dependent
magneto-resistance (ADMR) \cite{Abdel-jawad07prl} or the Raman
scattering experiments.\cite{Devereaux07rmp} For instance, the
polar ADMR measurements in OD Tl$_2$Ba$_2$CuO$_{6+\delta}$
(Tl2201) were analyzed in terms of the transport scattering rate
$\Gamma_{tr}$ which consists of the isotropic and anisotropic
terms as
 \ba
 \label{gammatr}
\Gamma_{tr}(\theta,T)=\Gamma_0 (\theta) +\Gamma_1 \sin^2(2\theta)
T +\Gamma_2 T^2,
 \ea
where $\Gamma_0 (\theta)$ is proportional to $1/v_F(\theta)$,
i.e., the in-plane density of states.\cite{Varma01prl} The
anisotropic $T$ linear term, interestingly, has the same
anisotropic form as the $\omega$ linear contribution to the $Im
\Sigma$ that Chang $et~al.$ reported, as mentioned above.

Besides its own significance, the momentum anisotropy of the
single-particle self-energy is also necessary to understand the
enormous amount of spectroscopic and transport properties of the
cuprates. Moreover this self-energy is determined by coupling to
the fluctuation spectra which is the essential physical quantity
to understand in the cuprates. This may be done using the
inversion of the Eliashberg equations provided the reliability of
the approximations in the latter is checked consistently.
Recently Schachinger and Carbotte\cite{Schachinger08prb}
performed this inversion using the maximum entropy method
(MEM)\cite{Shi04prl} from the nodal cut ARPES data on Bi2212 of
Zhang $et~al.$\cite{Zhang08prl} The obtained fluctuation spectra
showed a weak peak around 0.06 eV and a cut-off near 0.4 eV,
which is similar to that obtained from the frequency dependent
conductivity on Bi2212.\cite{Hwang07prb} This may be anticipated
because the conductivity, being proportional to the Fermi
velocity squared, is strongly weighted by contributions from near
the nodal region. The analysis of the frequency dependent
conductivity in terms of the self-energy and Eliashberg function
has been applied in many
cases.\cite{Gimm02self,Choi03ijmpb,Dordevic09prb,Heumen09prb}

\begin{figure}
\includegraphics[scale=0.4]{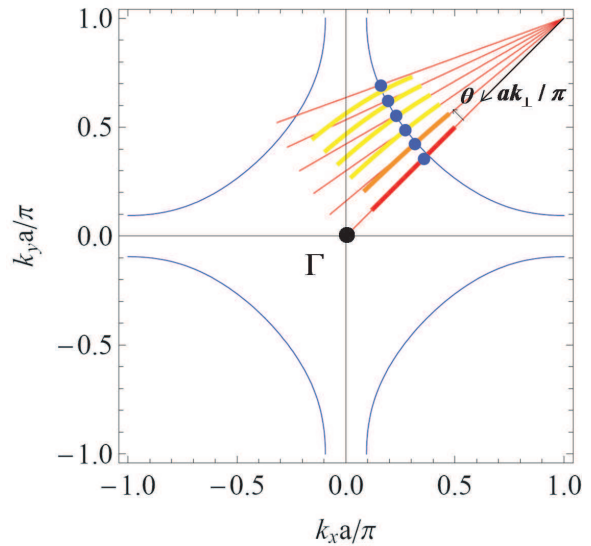}
\caption{The Fermi surface of Bi2212 in the first Brillouin zone.
The blue solid curve is the FS from Eq.\ (\ref{eq:tight}) and the
solid dots are the experimentally determined FS at $\theta=0$, 5,
10, 15, 20, and 25 degrees. $k_\perp$ is the distance from the
$(\pi,\pi)$ point. The thick bars along each cut indicate the
ranges of experimentally measured ARPES MDC data. Parts of them
which do not deviate much from the lines were actually used in the
TB fitting as indicated in Fig.\ \ref{fig:MDCfit}.} \label{fig:FS}
\end{figure}

However, the information obtained from the nodal direction alone
is insufficient to characterize the self-energy and the momentum
dependence of the fluctuation spectra. In the present work we
analyze the momentum anisotropy and frequency dependence of the
qp self-energy $\Sigma(\theta,\omega)$ in a slightly under-doped
Bi2212 sample of $T_c=89$ K along the cuts of the tilt angles
$\theta=$ 0, 5, 10, 15, 20, and 25 degrees with respect to the
nodal direction from $(\pi,\pi)$ point in the Brillouin zone as
shown in Fig.\ \ref{fig:FS}. The angle-resolved photoemission
measurements have been carried out on a VUV laser-based ARPES
system.\cite{Liu08rsi} In the experimental setup, there are two
angles, $\theta$ and $\phi$, to be controlled. They are then
converted to the actual tilt angle $\tilde\theta$ and the
amplitude $k_{\perp}$ of the inplane wave-vector $\bf k$. The
$k_{\perp}$ in this paper is referred to the distance from the
$(\pi,\pi)$ point. The thick bars along the cuts of the Fig.\
\ref{fig:FS} indicate the range of $k_\perp$ of the collected
ARPES data as the angle $\phi$ is varied by approximately 30
degrees. The actual tilt angle $\tilde\theta$ deviates slightly
from the control tilde angle $\theta$ as the $\phi$ is varied.
The change of $\theta$ is small and is disregarded in the
analysis to be presented here. See the more technical discussion
below Eq.\ (\ref{eq:LD}) in section II.

\begin{figure}
\includegraphics[scale=0.7]{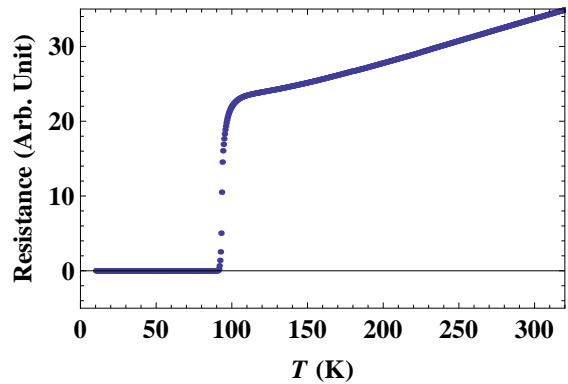}
\caption{The temerature dependence of the resistivity of a
slightly underdoped Bi2212 of $T_c=89$ K. $T^*$ is estimated from
the deviation from the linearity to be $ T^* \approx 160$ K.}
\label{fig:res}
\end{figure}

The photon energy of the laser is 6.994 eV with a bandwidth of
0.26 meV. The energy resolution of the electron energy analyzer
(Scienta R4000) is set at 1 meV, giving rise to an overall energy
resolution of 1.03 meV which is significantly improved from
regular 10$\sim$15 meV  from regular synchrotron radiation
systems. The angular resolution is $\sim 0.3^\circ $,
corresponding to a momentum resolution $\sim$0.004 \AA$^{-1}$ at
the photon energy of 6.994 eV, more than twice improved from
0.009 \AA$^{-1}$ at a regular photon energy of 21.2 eV for the
same angular resolution. The slightly-underdoped Bi2212 single
crystals with a superconducting transition temperature $T_c$=89 K
were cleaved {\it in situ} in vacuum with a base pressure better
than 5$\times$10$^{-11}$ Torr. The measurements were carried out
at $T=107$ K which is below the pseudogap temperature $T^*$. The
pseudogap temperature of the sample may be determined from the
temperature dependence of the resistivity shown in Fig.\
\ref{fig:res}. From the deviation from the $T$ linearity of the
resistivity, it is approximately $ T^* \approx 160$ K in
agreement with the independent estimate from different
group.\cite{Kanigel06naturephys}

By the Lorentzian fitting of the ARPES momentum distribution
curve (MDC) shown in Fig.\ \ref{fig:MDCfit}, we extract the
self-energy $\Sigma(\theta,\omega)$ (shown in Fig.\
\ref{fig:sigma107} below) using a realistic tight-binding band
dispersion (Eq.\ (\ref{eq:tight}) below). We confirm that the
normal state self-energy exhibits an in-plane momentum anisotropy
in both the elastic and inelastic parts. Details will be
presented in Section II. From the obtained self-energies, we
extract the  function $\alpha^2 F$ by inverting the Eliashberg
equation. We employ the adaptive maximum entropy method. Its
formulation is given in Section III. We show that in spite of the
anisotropy in $\Sigma(\theta,\omega)$ the $\alpha^2
F(\theta,\omega)$ obtained from data from the cuts at different
$\theta$'s collapse onto a single curve (Fig.\
\ref{fig:Efunc107}) at low energies with a cut-off at about 0.4
eV around the nodal direction changing smoothly to approximately
0.2 eV at 25 degrees. The variation of the cut-off is simply
accounted for by the variation in the position of the bottom of
the band with respect to the Fermi-energy with $\theta$ with the
intrinsic cut-off of the spectrum of about 0.4 eV.
Some of these results were anticipated by an
approximate calculation\cite{Zhu08prl} using an assumed spectra
by analysis of ARPES in La$_{2-x}$Sr$_x $CuO$_4$, rather than the
more reliable inversion method used here. These results put strong
constraints on the momentum dependence of the fluctuation
spectra, which we will discuss later. An implication of our
results is that the anisotropy of the transport and spectroscopic
properties originate simply from the anisotropy of the Fermi
surface. The detailed results are presented in Section IV. In
Section V, we will conclude with  a discussion of the results,
summary and remarks motivating further work.

\section{Deducing the Self-energy}

The ARPES intensity, within the sudden approximation, is given by
 \ba
I(\vk,\omega)= |M(\vk,\nu)|^2 f(\omega)\left[
A(\vk,\omega)+B(\vk,\omega) \right],
 \ea
where $M(\vk,\nu)$ is the matrix element, $\nu$ the energy of
incident photon, $f(\omega)$ the Fermi distribution function, $A$
the qp spectral function, and $B$ is the background. We write the
in-plane momentum $\vk$ with the $k_\perp$ perpendicular to the
Fermi surface (FS) and the angle $\theta$ measured from the nodal
cut as indicated in Fig.\ \ref{fig:FS}. Since $\Sigma$ has a much
weaker dependence on $k_\perp$ than $\omega$ (which is verified
in the experiments through the Lorentzian distribution of the
spectra as a function of $(k_\perp-k_F(\theta))$, the spectral
function may be rewritten as
 \beeq
 \label{eq:spectral}
A(\theta,k_\perp,\omega)=-\frac{1}{\pi}
\frac{\Sigma_{2}(\theta,\omega)}{[\omega-\xi(\mathbf{k})-\Sigma_{1}(\theta,\omega)]^2
 +[\Sigma_{2}(\theta,\omega)]^2}.
 \eneq
Here, $\xi(\mathbf{k})$ is the bare dispersion, and $\Sigma_{1}$
and $\Sigma_{2}$ are, respectively, the real and imaginary parts
of the self-energy. Along a cut with a fixed tilt angle $\theta$,
then, the dependence on $k_\perp$ is solely through the bare
dispersion energy $\xi(\vk)$. $A$ has a Lorentzian form with
respect to $\xi(\vk)$. The
self-energy can be directly extracted from the ARPES intensity by
a Lorentzian fit if $\xi(\vk)$ is known.

\begin{figure*}
\includegraphics[scale=0.7]{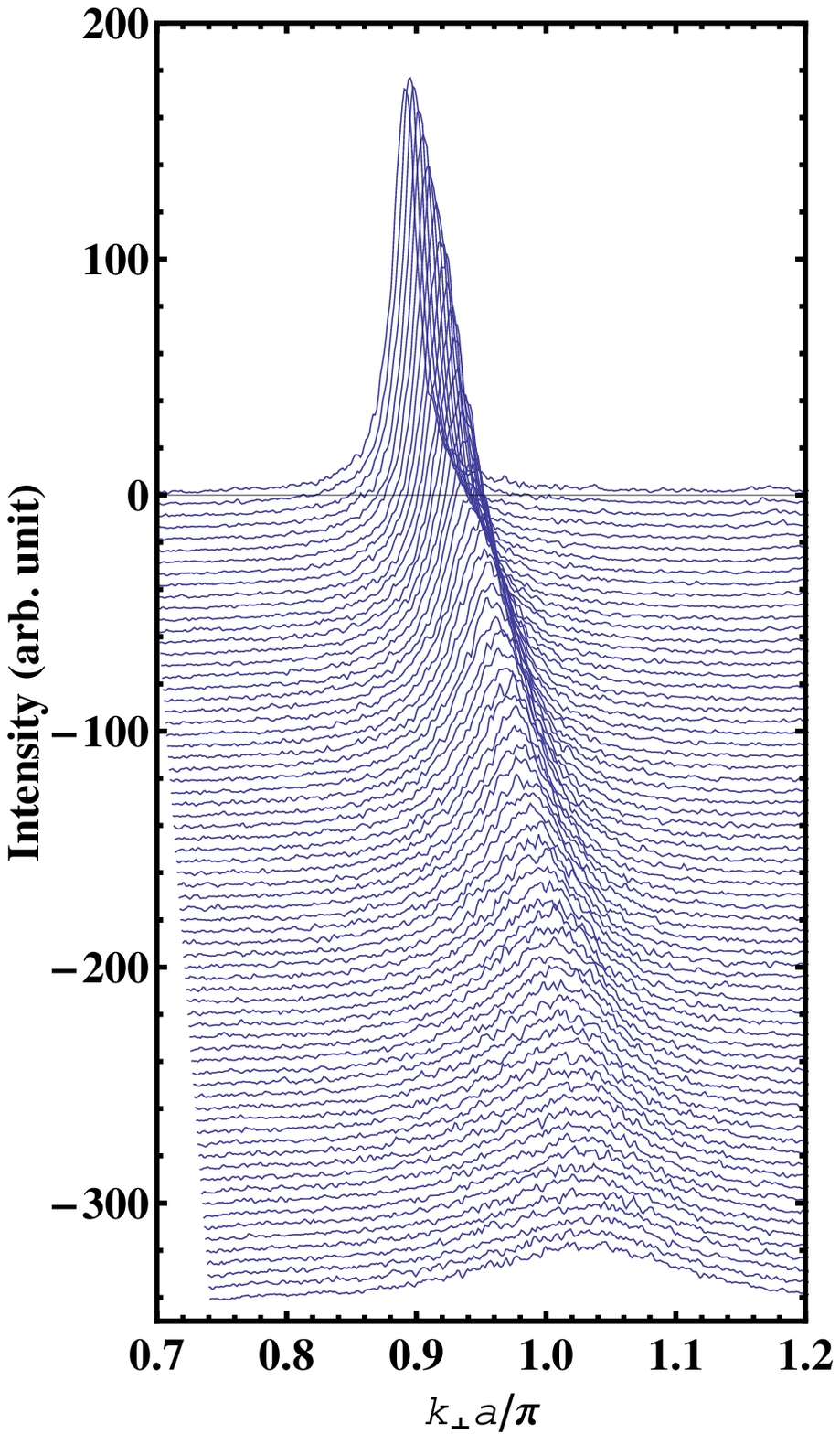}
\includegraphics[scale=0.7]{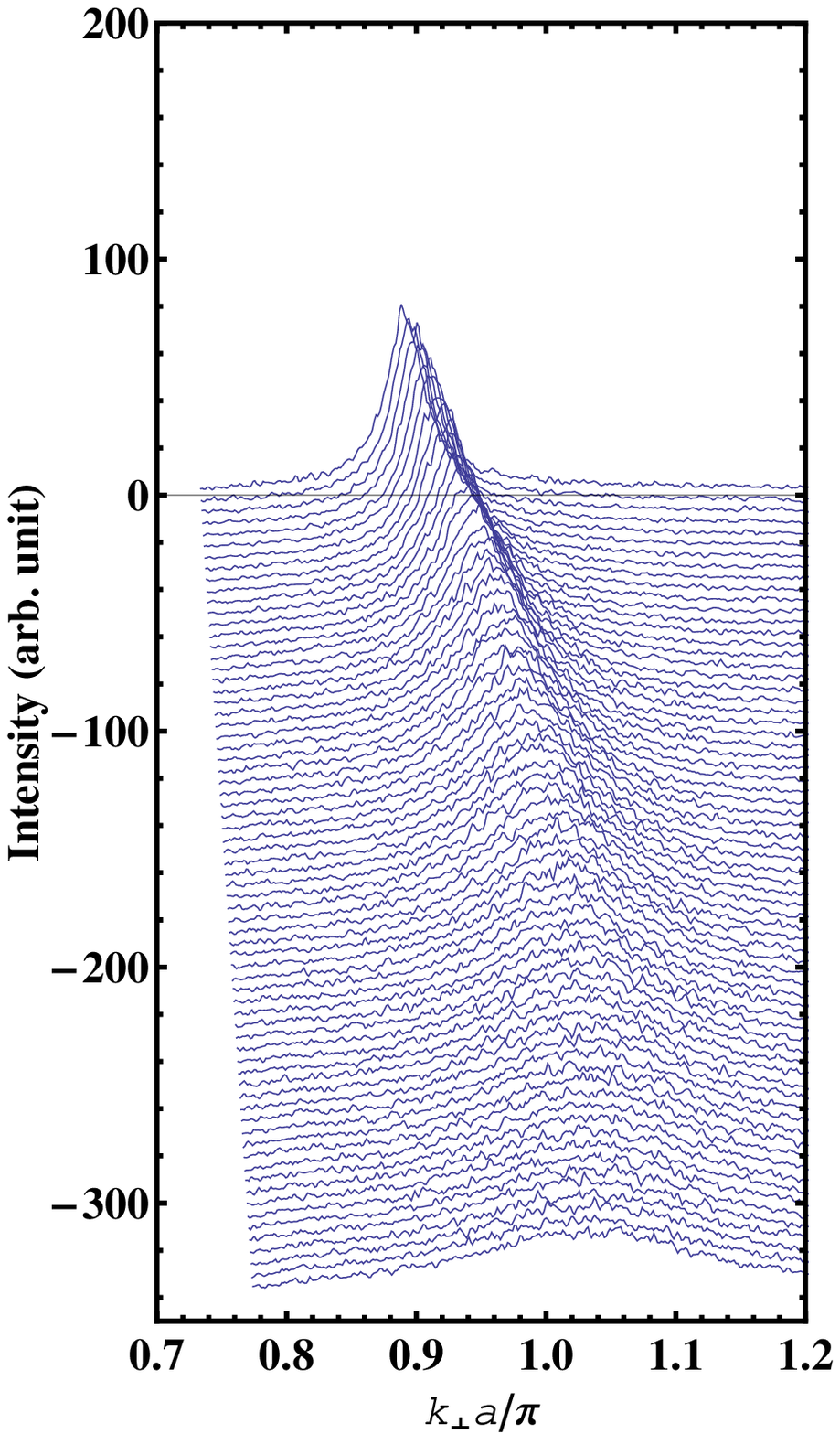}
\caption{The representative momentum distribution curves (MDC) as
a function of the momentum perpendicular to the FS, $k_\perp
a/\pi$, and their dispersion as the binding energy is varied. Each
curve is shifted down for clarity. The left plot is for the tilt
angle 5 and the right for 20 degrees. From the top to bottom are
the MDC at the energy $\omega=0.0005$ eV with the step of 0.005
eV up to 0.3455 eV. The qp coherence peak becomes suppressed as
the energy is increased away from the Fermi level or the tilt
angle increased from the nodal cut. } \label{fig:MDCdisp}
\end{figure*}

We first determined the tight-binding (TB) dispersion of
$\xi(\vk)$ by matching the experimental Fermi surface with the
four parameter dispersion.
 \begin{eqnarray}\label{eq:tight}
\xi(k_{x},k_{y}) = -2t(\textrm{cos} k_{x}a+\textrm{cos}
k_{y}a)+4t'\textrm{cos} k_{x}a \textrm{cos} k_{y}a
 \nonumber \\
-2t''(\textrm{cos}2k_{x}a+\textrm{cos}2k_{y}a)-\mu ,
 \end{eqnarray}
where $a=3.82$ \AA \ is the lattice constant and $\mu$ is the
chemical potential. We took $t=0.395$, $t'=0.084$, $t''=0.042$,
and $\mu=-0.43$ eV, which are consistent with Kordyuk
$et~al$.\cite{Kordyuk03prb} Note that we neglected the bilayer
splitting present in the Bi2212 compounds. At the photon energy
used ($h\nu=6.994$ eV) in the laser ARPES, only the antibonding
bands are observed and the bonding band is completely suppressed.
The experimentally determined FS in comparison with that from Eq.\
(\ref{eq:tight}) is shown in Fig.\ \ref{fig:FS}. The 6 cuts with
the tilt angles $\theta$ with respect to the $(\pi,\pi)$ are also
shown with the solid lines. To study the importance of the bare
dispersion $\xi$ we also used the linear dispersion (LD) for
comparison.
 \beeq \label{eq:LD} \xi(\vk)=v_{F}(\theta) \left[ k_\perp-k_{F}(\theta) \right],
 \eneq
where $v_{F}$ and $k_{F}$ were calculated from the tight binding
dispersion of Eq.\ (\ref{eq:tight}).

In order are some technical points about the cuts perpendicular
to the Fermi surface. In the ARPES experiment setup, two angles
are the control parameters: one is the tilt angle $\theta$ which
closely corresponds to the actual tilt angle $\tilde\theta$ and
the other is $\phi$ which determine the $k_\perp$ such that
 \ba
k_0 = \frac{\sqrt{2mE_{kin}}}{\hbar},~~ k_x = k_0 \sin\phi,~~
 k_y= k_0 \cos\phi \sin\theta,
 \ea
where $E_{kin}$ is the kinetic energy of the photoelectons given
by
 \ba
E_{kin}=h\nu-W-|\omega|,
 \ea
where $W=4.3$ eV is the work function. The $k_x$ and $k_y$ are the
components of the wave-vector of the photoelectron with respect
to the diagonal cuts. The $k_\perp$, which is the distance from
the $(\pi,\pi)$ point can be simply calculated by the
trigonometry. The $\phi$ is varied by approximately 30 degrees
and the corresponding $k_\perp$ which go through the Fermi
surface is shown by the thick bars in Fig.\ \ref{fig:FS}. A
consequence of this experimental setup is that the actual tilt
angle $\tilde\theta$ is not constant as given by the $\theta$.
$\tilde\theta=\theta$ for $\theta=0$, but $\tilde\theta$ begins
to deviate from $\theta$ as $\phi$ is varied when $\theta\ne 0$.
The actual cuts therefore are not straight lines but slightly
curved for large $k_\perp$ as shown in Fig.\ \ref{fig:FS}, but
for $k_\perp$ near the Fermi surface the cuts point to the
$(\pi,\pi)$ direction. We disregard the difference between the
$\tilde\theta$ and $\theta$, and use $\theta$ to denote the tilt
angle in this paper.

Typical ARPES intensity as a function of $k_\perp$ for fixed tilt
angles and binding energy, referred to as the momentum
distribution curve, and the dispersion as a function of the
binding energy $\omega$ are shown in Fig.\ \ref{fig:MDCdisp}. Each
curve is shifted down for clarity as the binding energy is varied.
The left and right plots are, respectively, for the tilt angle 5
and 20 degrees. From the top to bottom are the MDC at the energy
$\omega=0.0005$ eV with the step of 0.005 eV up to 0.3455 eV.
Note that the qp coherence peak becomes suppressed as the energy
is increased away from the Fermi level or the tilt angle
increased from the nodal cut. These MDC were fitted by equating
the ARPES intensity with the spectral function given by Eq.\
(\ref{eq:spectral}) as
 \ba
I(\theta,k_\perp,\omega)= C
A(\theta,k_\perp,\omega)+B(\theta,\omega)
 \ea
to extract the self-energy as a function of the energy for a
given cut.

\begin{figure*}
\includegraphics[scale=1.]{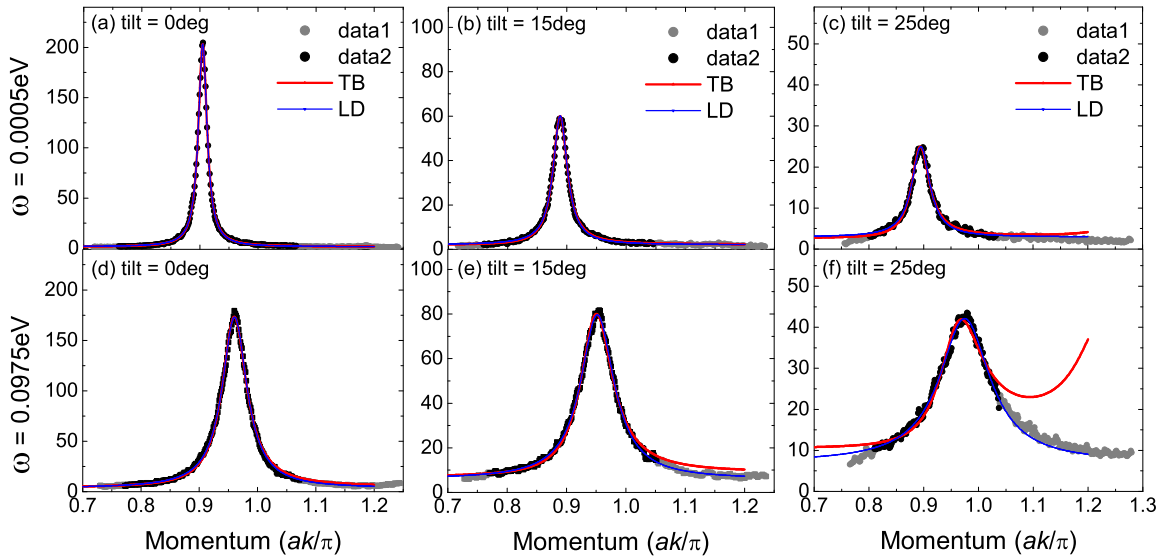}
\caption{The representative momentum distribution curves as a
function of the momentum perpendicular to the FS, $k_\perp
a/\pi$, at the tilt angles $\theta=0$, 15, and 25 degrees, and the
binding energy $\omega=0.0005$ and $0.0975$ eV. The shaded
circles are the experimental ARPES intensities and the solids
ones are those used in the fitting. The red and blue lines are
the fitting with the TB and LD bands, respectively. Note that the
MDC is not Lorentzian when plotted as a function of $k_\perp$
with the TB dispersion as can clearly be seen from (f). }
\label{fig:MDCfit}
\end{figure*}

In Fig.\ \ref{fig:MDCfit}, we show the MDC fits for the nodal
($\theta=0$), $\theta=15$, and $\theta=25$ degrees cuts and for
the binding energy $\omega=0.0005$ (the Fermi level) and
$\omega=0.0975$ eV as representative cases. The shaded dots are
the experimental ARPES intensities and the solid dots are those
used in the fitting. The red and blue lines are the spectral
function of Eq.\ (\ref{eq:spectral}) with the TB and LD bands,
respectively. From the peak position and the width of the peak
together with the bare dispersion $\xi(\vk)$, we can determine
the real and imaginary parts of the self-energy.

\begin{figure}
\centering
\includegraphics[scale=1.]{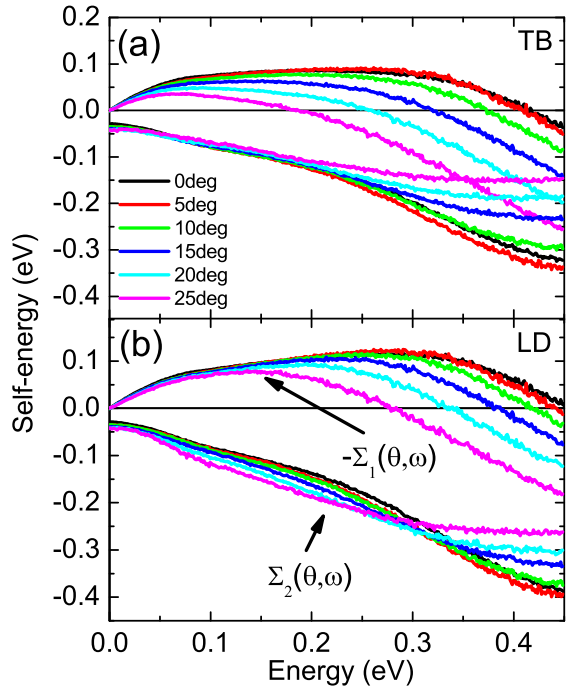}
\caption{The real and imaginary part self-energies at $T=107$ K
for the tilt angles $\theta=0$, 5, 10, 15, 20, and 25 degrees.
$-\Sigma_1$ and $\Sigma_2$ as a function of the positive energy
$\omega$ are shown. Plot (a) shows the extracted self-energy
using the TB dispersion, and (b) shows, for comparison, one using
the linear dispersion. }
 \label{fig:sigma107}
\end{figure}

The self-energies, $-\Sigma_1(\theta,\omega)$ and
$\Sigma_2(\theta,\omega)$ for $\omega>0$, determined this way at
$T=107$ K are shown in Fig.\ \ref{fig:sigma107}. The plot (a) was
obtained using the TB and the plot (b) using the LD band. The real
parts of the self-energy, $\Sigma_1(\omega)$, cross the zero at
progressively smaller energies as the tilt angle is increased.
This feature is more pronounced in the TB dispersion analysis as
can be seen by comparing the plots (a) and (b), and is better
described by the TB because the band bottom can not be captured
by the LD. The imaginary parts of the self-energy decrease
monotonically as $\omega$ is increased up to $\omega=0.45$ eV. The
elastic part, $\Sigma_2 (\theta,\omega=0)$, clearly changes as
$\theta$ is changed. The elastic qp scattering rate is momentum
anisotropic in accord with previous
works.\cite{Kaminski05prb,Chang08prb,Abdel-jawad07prl} The
functional dependence of the qp scattering rates on the $\theta$
and $\omega$ for small $\omega$ may be analyzed in analogy with
the Eq.\ (\ref{gammatr}). We confirm that both elastic and
inelastic qp scattering rates exhibit an anisotropy as a function
of angle $\theta$ around the Fermi-surface.

The extracted self-energies were used as an input to dudece the
fluctuation spectral functions by inverting the Eliashberg
equation. It will be presented in the following section. Before
the detailed analysis, we note that the relation
 \ba
-\frac{\partial\Sigma_2(\theta,\omega)}{\partial\omega} =\pi
\alpha^2 F(\theta,\omega) \label{eq:dsigma2}
 \ea
holds for $\omega\gg T$, as may be deduced from Eq.\
(\ref{eq:imself}) below. The extracted $\Sigma_2(\theta,\omega)$
from the TB collapse onto a single curve up to $\omega \lesssim
0.2$ eV as can be seen from Fig.\ \ref{fig:sigma107}. This,
together with Eq.\ (\ref{eq:dsigma2}), suggests that the
Eliashberg function $\alpha^2 F(\theta,\omega)$ would yield a
single curve independent of the tilt angle $\theta$ for $T\ll
\omega \lesssim 0.2$ eV. The detailed analysis presented in the
following sections will establish that the Eliashberg function
$\alpha^2 F(\theta,\omega)$ at different angle $\theta$ collapse
onto a single function of $\omega$ with the intrinsic cut-off of
about 0.4 eV or the energy of the band bottom in direction
$\theta$ with respect to the Fermi energy, whichever is smaller.
This is our key result which we will turn to.

\section{Deducing the Fluctuation Spectra: Formulation}

Neglecting vertex corrections (justification for this will be
given in the concluding section), the self-energy
$\Sigma(\vk,\omega)$ may be written as
 \ba
\label{eq:Eliashberg01}
 \Sigma(\vk,\omega)=\int_{-\infty}^{\infty}
d\epsilon \int_{-\infty}^{\infty} d\epsilon'
\frac{f(\epsilon)+n(-\epsilon')}{\epsilon+\epsilon' - \omega -
\imath \delta} \nonumber \\
\times \sum_{\vk'} A(\vk',\epsilon) \alpha^2(\vk,\vk')
F(\vk',\vk,\epsilon'),
 \ea
where $f$ and $n$ represent the Fermi and Bose distribution
function, respectively. $\alpha({\bf k,k'})$ is the matrix element
for scattering fermions with the fluctuations of spectral weight
$F(\vk',\vk,\epsilon)$. After the integral over $k'_\perp$, the
real and imaginary parts are given by
 \ba \label{eq:reself}
\Sigma_{1}(\theta,\omega)=\int_{-\infty}^{\infty} d\epsilon'
\int_{-\infty}^{\infty} d\epsilon
\frac{f(\epsilon)+n(-\epsilon')}{\epsilon+\epsilon' - \omega}
\alpha^2 F(\theta,\epsilon'), \\
\Sigma_{2}(\theta,\omega)=\pi\int_{-\infty}^{\infty}d\epsilon'
\left[ f(\omega-\epsilon')+n(-\epsilon') \right] \alpha^2
F(\theta,\epsilon') ,
 \label{eq:imself}
 \ea
where the Eliashberg function, or the bosonic coupling spectrum,
$\alpha^2 F(\theta,\epsilon')$ is given by
 \ba
 \label{eq:alpha2F}
\alpha^2 F(\theta,\epsilon')\equiv \left\langle
\frac{\alpha^2(\theta,\theta')}{v_F(\theta')}
F(\theta,\theta',\epsilon') \right\rangle_{\theta'} .
 \ea
The $v_F(\theta')$ is the angle dependent Fermi velocity and the
bracket implies the angular average. We can extract the
``averaged'' spectrum $\alpha^2 F(\theta,\epsilon')$ by inverting
either Eqs.\ (\ref{eq:reself}) or (\ref{eq:imself}). Thus
information only about the angle dependence of the average over
the product of the squared matrix element and spectra as
constrained by energy-momentum conservation are obtained. Both
equations must give the same results for $\alpha^2
F(\theta,\omega)$ provided the real and imaginary parts of the
self-energy satisfy the Kramers-Kronig (KK) relation. In the
present case, however, the extracted spectra do depend on whether
it is extracted from the real or imaginary part. It implies that
the extracted real and imaginary parts of the self-energy do not
satisfy the KK to the required accuracy.
%
%
It turned out that the $\alpha^2F$ from the real part self-energy
is more reliable as shown below, probably because the peak
position of MDC is better determined than the width. This point
was also noted by Shi $et~al.$ \cite{Shi04prl}

A few techniques have been devised to invert the
Eliashberg equation, (\ref{eq:reself}) or (\ref{eq:imself}). We
employ the maximum entropy method (MEM).\cite{Shi04prl,numrecipe}
It is a useful technique to overcome the numerical instability in
the direct inversion, by incorporating the physical constraints
into the fitting process. The MEM minimizes the functional:
 \beeq \label{eq:L} L=\frac{\chi^{2}}{2}-\alpha S.
 \eneq
The $\chi^2$ is the error and $S$ is the generalized
Shannon-Jaynes entropy defined below. The multiplier $\alpha$ is
a determinative parameter that controls how close the fitting
should follow the data while not violating the physical
constraints. When $\alpha$ is small, the fitting will follow the
data as closely as possible at the expense of a noisy and/or
negative Eliashberg function, and when $\alpha$ is large, the
extracted Eliashberg function will not deviate much from the
constraint function $m(\epsilon')$. For a given tilt angle
$\theta$, we take
 \ba
\chi^2 = \sum_{i=1}^{N_D} \frac{\left[ D_i - \Sigma_1 (\omega_i)
\right]^2}{\sigma_i^2},\qquad\qquad\qquad\qquad\qquad\qquad\,\\
S=\int_{0}^{\infty}d\epsilon' \left[ \alpha^2
F(\epsilon')-m(\epsilon')-\alpha^2
F(\epsilon')\textrm{ln}\frac{\alpha^2 F(\epsilon')}{m(\epsilon')}
\right],
 \ea
where $N_D$ is the total number of data points, $D_i$ is the
experimental data for the real part of the self-energy at the
energy $\omega_i$, $\Sigma_1(\theta,\omega_i)$ is defined by Eq.\
(\ref{eq:reself}), and $\sigma_i$ is the error bars of the data.
The entropy term imposes physical constraints to the fitting and
is maximized when $\alpha^2 F(\epsilon')=m(\epsilon')$, where
$m(\epsilon')$ is the constraint function. To minimize Eq.\
(\ref{eq:L}) for a given $m(\epsilon')$, the Eliashberg function
$\alpha^2 F(\omega_j)$ is optimized in each iterative step by
updating it as
 \ba
\delta F_j = -\sum_k A_{jk}^{-1} \frac{\delta L}{\delta F_k},
~~A_{jk}=\frac{\delta^2 L}{\delta F_j \delta F_k},
 \ea
where $F_j =\alpha^2 F(\omega_j)$. The matrix inversion of
$A_{jk}^{-1}$ was performed using the singular value
decomposition technique. Details of the algorithms can be found in
Ref.\ \cite{numrecipe}.

\begin{figure}
\includegraphics[scale=0.8]{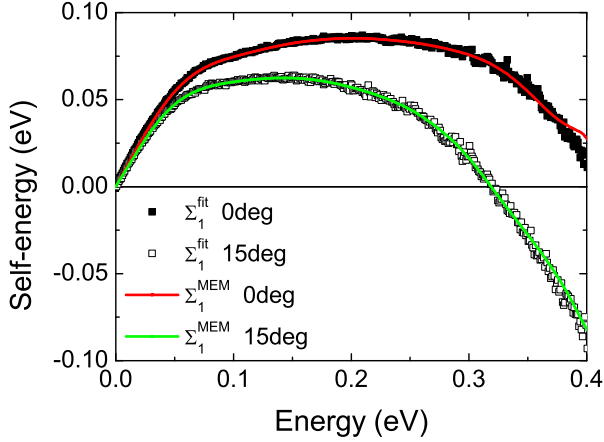}
\caption{Comparison between the real part of the self-energy and
the MEM fitting of Eq.\ (\ref{eq:reself}) for the tilt angles
$\theta=0$ and 15 degrees. The squares are the extracted real
part self-energy of Fig.\ \ref{fig:sigma107}(a), and the solid
lines are the MEM fitting.} \label{fig:compare}
\end{figure}

In usual MEM approaches, the constraint function is chosen based
on our best $a ~priori$ knowledge for the specific system and
remains unaltered. The obtained results for $\alpha^2
F(\epsilon')$ do depend on how the $m$ was chosen. In order to
decrease the dependence on the constraint function and also to
better represent the physical system, we employ the adaptive MEM
which was implemented as follows: For a chosen constraint
function $m(\epsilon')$, the iterative minimization is performed
as described above. After the convergence, the $m(\epsilon')$ is
updated to a linear combination of $m(\epsilon')$ and $\alpha^2
F(\epsilon')$ to reflect the nature of the solution. For this new
constraint function, the minimization is performed again via
iterations. This second optimization is repeated until the
convergence is reached. The double iterative adaptive MEM
decreases the dependence on the choice of the constraint function
and improves the overall fitting quality. The quality of the
adaptive MEM can be seen by comparing the real part self-energy
from the ARPES data and the self-energy expression of Eq.\
(\ref{eq:reself}) as shown in Figs.\ \ref{fig:compare}.

\section{Fluctuations Spectra}

\begin{figure}
\includegraphics[scale=1.1]{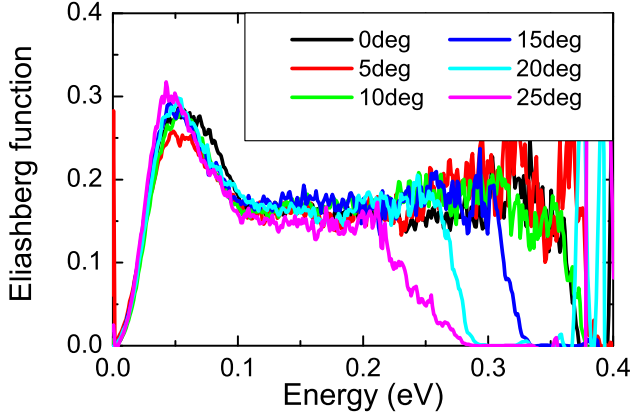}
\caption{ The Eliashberg function extracted from the real part of
the self-energy at $T=107$ K. Notice the remarkable collapse of
the Eliashberg functions below $\omega\approx 0.2$ eV for
different tilt angles. } \label{fig:Efunc107}
\end{figure}

\begin{figure}
\includegraphics[scale=0.4]{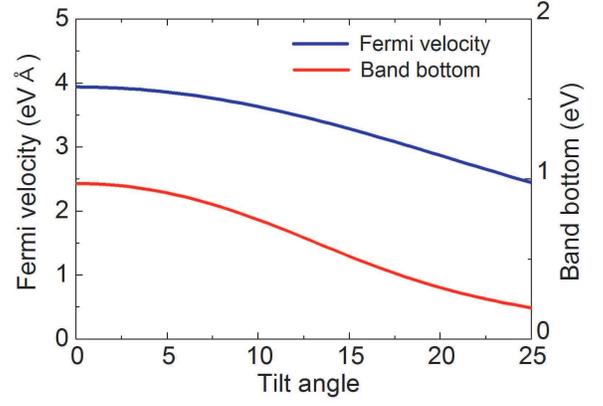}
\caption{The angle dependence of the Fermi velocity and band
bottom calculated from Eq.\ (\ref{eq:tight}).} \label{fig:bbottom}
\end{figure}

As noted previously, the real part of the self-energy is
determined more accurately from the ARPES intensity, and as the
tilt angle is varied the tight-binding dispersion is more
reliable. We therefore present in Fig.\ \ref{fig:Efunc107} the
Eliashberg function $\alpha^2 F(\theta,\omega)$ extracted from
the real part self-energy determined using the TB dispersion
shown in Fig.\ \ref{fig:sigma107}. Note that structures in
$\Sigma(\theta,\omega)$ are reflected in the coupled fluctuation
spectra $\alpha^2 F(\theta,\omega)$. The Eliashberg functions
increase approximately  linearly from zero as $\omega$ is
increased and have a peak at $\omega\approx 0.05$ eV, flattens
above 0.1 eV, and vanishes above a cut-off energy $\omega_c$
which does depend on $\theta$. The noisy feature above 0.35 eV is
the numerical artifacts of the singular value decomposition and
MEM. The cut-off energy is approximately 0.37 eV along the nodal
cut. These are in good agreement with Schachinger and
Carbotte\cite{Schachinger08prb} who analyzed the nodal cut Bi2212
laser ARPES data for the same sample.\cite{Zhang08prl}

Now we discuss the new results of the analysis of the off-nodal
cuts. The Eliashberg functions obtained are presented in Fig.\
\ref{fig:Efunc107} for the tilt angles $\theta=$ 0, 5, 10, 15,
20, and 25 degrees. As the tilt angle is increased, the peak
energy remains unaltered, but the cut-off energy decreases
monotonically. At $\theta=25$ degrees, $\omega_c\approx 0.25$ eV.
With Laser ARPES the data at large angles is not obtainable due
to kinetic constraints; moreover the problems presented in the
analysis due to interference of data due to the contributions
from the next zone, as is obvious from Fig.\ \ref{fig:FS}, become
more serious. An important point may be noted from comparison of
Fig.\ \ref{fig:Efunc107} and Fig.\ \ref{fig:bbottom} where the
position of the band bottom with respect to the Fermi-energy is
given. The latter is about 1 eV in the nodal direction and about
0.25 eV at $\theta  \approx$ 25 degrees. As was shown in a simple
calculation\cite{Zhu08prl} and as is natural, the band-bottom
serves as an effective cut-off in the fluctuation spectrum when
it is lower than the intrinsic cut-off. So one important results
is that the intrinsic cut-off of the spectrum is about 0.4 eV. An
even more important result contained in Fig.\ \ref{fig:Efunc107}
is that quite remarkably the fluctuation spectra for different
tilt angles collapse onto a single curve below the angle
dependent {\it effective} cut-off. Note that the corresponding
self-energies do change even at low energies as the angle is
varied as can be seen from Fig.\ \ref{fig:sigma107}. From Eq.
(\ref{eq:reself}) it follows that a nearly isotropic
$\alpha^2F(\theta, \omega)$ produces an angle dependent
self-energy  purely from the effects of the dispersion of the
bare band.

\begin{figure}
\includegraphics[scale=0.8]{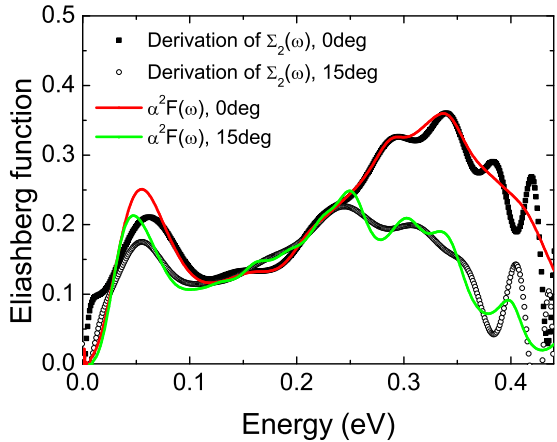}
\caption{The Eliashberg function deduced from the imaginary parts
of the self-energy. Results for $\theta=0$ and 15 degrees are
given for representative cases. Also shown are $-\frac1\pi
\frac{\partial\Sigma_2(\omega)}{\partial\omega}$. They agree
overall as expected. } \label{fig:Efrom2}
\end{figure}

The Eliashberg functions may also be extracted from the imaginary
parts of the self-energy using Eq.\ (\ref{eq:imself}). The
results are given in Fig.\ \ref{fig:Efrom2} for the tilt angles 0
and 15 degrees for representative cases. As alluded before, they
do not agree exactly with those from the real parts of the
self-energy. Especially the features around the cut-off are
exaggerated in the $\alpha^2 F(\theta,\omega)$ from the imaginary
part self-energies. The $\alpha^2F(\theta,\omega)$ do not
collapse as neatly as those from the real parts. Also shown are
the $-\frac1\pi \frac{\partial\Sigma_2(\omega)}{\partial\omega}$
for the same tilt angles. The $-\frac1\pi
\frac{\partial\Sigma_2(\omega)}{\partial\omega}$ and $\alpha^2
F(\omega)$ determined from $\Sigma_2(\omega)$ agree overall as
expected.

\section{Discussion of Results, Conclusions and Guides to Further Experiments}

\begin{figure}
\includegraphics[scale=0.7]{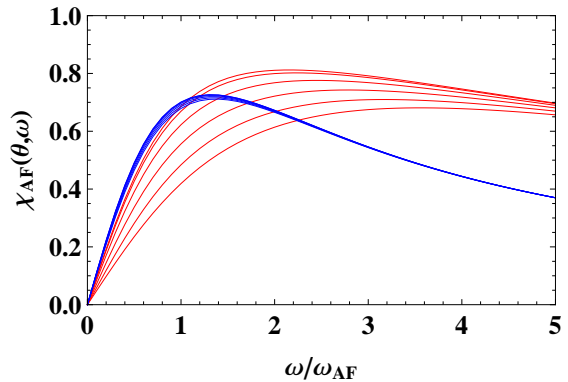}
\caption{The model Eliashberg function calculated from the
overdamped AF fluctuations of Eq.\ (\ref{eq:chiaf}). The red and
blue lines are for the AF correlation length $\xi/a=1$ and
$1/\pi$, respectively. For each $\xi$, the angles are from above
$\theta=0$, 5, 10, 15, 20, and 25 degrees. As $\xi$ is decreased
the fluctuation spectra become momentum isotropic. }
\label{fig:chiaf}
\end{figure}

In this section, we discuss the implications of our findings
which point to future experiments and analysis. They are the
angle independence of the Eliashberg function $\alpha^2F(\theta,
\omega)$, i.e the angle averaged product of the matrix element
and the spectral function of the fluctuations, as defined in Eq.\
(\ref{eq:alpha2F}), the possible physics of the low energy bump
around $\approx 0.05$ eV, the vertex corrections to the
Eliashberg equation, and various other assorted issues. The
finding that the Eliashberg function is angle independent below
the cut-off $\omega_c$ puts an important constraint on the
microscopic understanding of the cuprates. For example, it can
put a limit on the correlation length $\xi$ for the commonly
assumed form of the antiferromagnetic (AF) fluctuations. A
phenomenological form of the overdamped AF fluctuations may be
written as\cite{Millis90prb,Chubukov03}
 \ba
\chi_{AF}({\vk},\vk',\omega)= \frac{\alpha\xi^2 ~
\omega/\omega_{AF}}{(({\vk-\vk'}-{\bf Q})^2\xi^2 +1
)^2+(\omega/\omega_{AF})^2}. \label{eq:chiaf}
 \ea
$\chi_{AF}(\theta,\omega)$ can be obtained after the integral over
$\theta'$ with both $\vk$ and $\vk'$ on the Fermi surface, that
is,
 \ba
\chi_{AF}(\theta,\omega)= \left<
\chi_{AF}(\vk,\vk',\omega)\right>_{\theta'},
 \ea
where $\bf k$ and $\bf k'$ have the angle $\theta$ and $\theta'$
with respect to the nodal cut, respectively. A straightforward
calculation reveals that a weak $\theta$ dependence of
$\chi_{AF}(\theta,\omega)$ means that $\xi/a \ll 1$ for $\omega
\lesssim \omega_{AF}$, where $\omega_{AF}$ is the characteristic
AF energy scale. This is shown in Fig.\ \ref{fig:chiaf} for
$\xi/a=1/\pi$ with the blue and for $\xi/a=1$ with the red lines.
For each $\xi$, the angles are $\theta=0$, 5, 10, 15, 20, and 25
degrees from above. As expected, $\chi_{AF}(\theta,\omega)$
becomes angle independent as $\xi$ is decreased. The collapse of
$\alpha^2 F(\theta,\omega)$ implies that if it is due to the AF
fluctuations, the AF correlation length should be $\xi/a \lesssim
1/\pi$ for Bi2212 at $T=107$. This is an upper limit since we
have not included the angle dependence of the matrix elements of
coupling of AF fluctuation to fermions which enters Eq.\
(\ref{eq:alpha2F}). A correlation length less than a lattice
constant is indicative either that there is negligible spectral
weight in AF fluctuations or that they couple to single-particle
excitations weakly enough to have no observable effect. This
implies among others that the antiferromagnetic fluctuations may
not be underlying physics of the deduced fluctuation spectrum.
This also means that the scenario of "hot" and "cold"- spots on
the Fermi-surface is inapplicable to the cuprates.

An angle-independent $\alpha^2F(\theta, \omega)$ is consistent
with the quantum-critical spectra hypothesized in the marginal
Fermi-liquid description of cuprates and recently derived
microscopically\cite{Aji07prl,Aji09prb} to be the spectra in the
quantum-critical region of the phase diagram of the cuprates due
to the quantum melting of the loop-current order
observed\cite{Fauque06prl,Mook08prb,Li08nature,Kaminski02nature}
in the underdoped region of the cuprates. There is one aspect of
the deduced $\alpha^2F(\theta, \omega)$ which is not given by the
theory. This is the low energy bump at the energy of
$\omega\approx 0.05$ eV. The presence of this bump may be seen
directly in the deduced Im $\Sigma(\theta, \omega)$ which is not
linear as a function of $\omega$ for low $\omega \gg T$. The
linearity may be expected on the basis of such theory and earlier
ARPES experiments which, however, do not have the high resolution
of the present experiments. If the bump occurs only in samples
studied in the pseudogap region, there must exist collective modes
special to this region of the phase diagram. This can be checked
by equally high resolution data in samples in the
quantum-critical region or the overdoped region of the phase
diagram.

We now discuss the validity of the neglect of vertex correction
in the relation of the self-energy and the fluctuations. The
exact relation between the vertex correction $\delta \Lambda
({\bf k,q}, \omega, \nu)$ and the self-energy is that
 \ba \delta
\Lambda ({\bf k,q}, \omega, \nu) = \Sigma({\bf k+q}, \omega+\nu)
- \Sigma({\bf k}, \omega).
 \ea
As discussed, the experiments give at energies up to about 0.2
eV, that the self-energy has a momentum dependence due only to
the anisotropy of the Fermi velocity, which from Fig.\
\ref{fig:bbottom} is only about 20 \%. This may be regarded as
negligible. It therefore follows that the momentum dependence of
the vertex correction is negligible. This means that all the
essential conclusions about the momentum independence of the
fluctuation spectra arrived at here remain valid. As regards the
frequency dependence, there is indeed a vertex correction which in
dimensionless form is of $O(\alpha^2\omega_c/W)$, where
$\omega_c$ is the upper-cutoff of the fluctuations and $W$ is the
electronic bandwidth of about 2 eV. For $\alpha^2 \approx 1$,
there are then vertex corrections of $O(1/5)$. This means that
our {\it quantitative} conclusions have a validity no better than
about 20 \%.

One final remark pertains to the information that the deduced
spectra offers for the fluctuations spectra responsible for the
superconductive instability. The normal state single-particle
self-energy has the full symmetry of the lattice. It then
follows, as is visible from Eq.\ (\ref{eq:Eliashberg01}) that the
``$\ell=2$'' part of the fluctuation spectrum, which determines
the symmetry of the superconductivity in cuprates is not visible
through study of the normal state. It is unlikely that a
completely different form of the spectra is responsible for the
normal state and the superconducting state. A test of this
conjecture and the deduction unambiguously of the spectra
responsible for superconductivity requires a study of the
variation of the single-particle self-energy as a function of
energy on going from the normal to the superconducting state. This
would be a generalization\cite{Vekhter03prl} of the
McMillan-Rowell procedure\cite{Mcmillan65prl} to $d$-wave
superconductors requiring high resolution ARPES. Such work is in
progress.

To summarize, we have presented the self-energy analysis from the
ultra high resolution laser ARPES on the slightly underdoped
Bi2212 high temperature superconductors. The self-energy was
determined along the nodal and off-nodal cuts in the normal
state. Both the elastic and inelastic quasi-particle scattering
rates exhibit in-plane momentum anisotropy. The deduced
self-energies were then used as an experimental input to invert
the Eliashberg equation to extract the product of the fluctuation
spectra and the coupling to the single particles. The adaptive MEM
was used. The high resolution ARPES data together with the
realistic tight-binding band dispersion enabled us to determine
high quality self-energy and the Eliashberg function. At the
temperature $T=107$ K between the superconducting $T_c$ and
pseudogap temperature $T^*$, the Eliashberg functions for
different tilt angles collapse onto a single curve up to the upper
cut-off despite the angle dependent self-energy. The cut-off has
an intrinsic value of about 0.4 eV or the band-bottom energy with
respect to the Fermi level if it is less than about 0.4 eV. The
implications of our results have also been discussed.

\begin{acknowledgments}

JMB, JHY, and HYC acknowledge the support from Korea Research
Foundation (KRF) through Grant No.\ KRF-C00241. XJZ thanks the
support by the NSFC, the MOST of China (973 project No:
2006CB601002, 2006CB921302). XJZ also thanks Sheng Bing and Prof.
Hai-hu Wen for their help in resistivity measurement. CMV's work
is partially supported by the award DMR-0906530 of the US
National Science Foundation.

\end{acknowledgments}

\bibliographystyle{apsrev}
\bibliography{ref2}

\end{document}